\documentclass[%
 reprint,
 amsmath,amssymb,
 aps,
]{revtex4-1}

\usepackage{graphicx}% Include figure files
\usepackage{graphics}
\usepackage{dcolumn}% Align table columns on decimal point
\usepackage{bm}% bold math
\usepackage{float}
\usepackage{caption}
\usepackage{subfigure}
\usepackage{amsmath}
\usepackage{mathrsfs}
\usepackage{amssymb}
\begin{document}

\preprint{}

\title{Slip length of confined liquid with small roughness of solid-liquid interfaces}

\author{Li Wan}
\email{lwan@wzu.edu.cn}
\author{Yunmi Huang}
\affiliation{Department of Physics, Wenzhou University, Wenzhou 325035, P. R. China}
\date{\today}% It is always \today, today,

\begin{abstract}
We have studied the slip length of confined liquid with small roughness of solid-liquid interfaces. Dyadic Green function and perturbation expansion have been applied to get the slip length quantitatively. In the slip length, both effects of the roughness of the interfaces and the chemical interaction between the liquid and the solid surface are involved. For the numerical calculation, Monte Carlo method has been used to simulate the rough interfaces and the physical quantities are obtained statistically over the interfaces. Results show that the total slip length of the system is linearly proportional to the slip length contributed from the chemical interaction. And the roughness of the interfaces plays its role as the proportionality factor. For the roughness, the variance of the roughness decreases the total slip length while the correlation length of the roughness can enhance the slip length dramatically to a saturation value. 
\begin{description}
\item[PACS numbers]
47.10.ad, 47.15.-x, 47.61.-k, 87.85.Rs,47.63.mf
\item[Keywords]
Slip length, confined liquid, rough interface, correlation length
\end{description}
\end{abstract}

%\pacs {66.70.-f,85.80.-g,65.40.gp,72.25.-b,63.21.-e,44.10.+i}% PACS, the Physics and Astronomy
\maketitle

%\tableofcontents
\section{Introduction}
The dynamics of liquid confined in nano structures have attracted interests due to their potential applications in nanofluidics ~\cite{Eijkel,Bocquet1,Karnik1,Karnik2,Schasfoort,Siwy}. It has been considered that the Navier-Stokes(NS) equation is still valid for the description of the fluid hydrodynamics even if the liquid is scaled down to a few molecular  layers~\cite{Chan,Becker,Bocquet2,Georges,Leng,Li,Maali,Raviv}. In order to solve the NS equation for the confined liquid, a proper boundary conditions(BC) should be applied at the solid-liquid interfaces. A commonly used BC is the no-slip BC, which requires that the fluid has zero velocity at the interfaces. Such BC increases the hydrodynamic resistance to the fluid when the size of the confined liquid is decreased, which brings difficulties for the applications of the nanofluidics~\cite{Bocquet2}. In order to overcome the difficulties, the slippage of liquid at the solid surfaces has been introduced to decrease the hydrodynamic resistance ~\cite{Bocquet1,Bocquet2}. The BC with the slippage of the liquid at the interfaces is known as the Navier boundary condition (NBC). The NBC requires that the velocity components of the liquid does not vanish at the interfaces in the moving direction. In order to quantify the slippage of liquid, a slip length $b$ has been introduced in the NBC ~\cite{Batchelor,Bocquet3,Pit,Cottin1,Cottin2,Thompson,Cieplak,Zhu,Priezjev,Jabbarzadeh}. Generally, the no-slip BC can be recovered from the NBC by setting $b=0$. And, larger the slip length $b$, more slippery the liquid at the interfaces. Since the slip length $b$ used in the NBC determines the solutions to the NS equation for the confined liquid, many authors have proposed various structures for the interfaces to enhance the slip length $b$ ~\cite{Liu, Ou1,Ou2,Choi1,Choi2,Callies,Truesdell,Bocquet4,Joseph,Lauga,Richardson,Cottin4}. But for the various structures, the fundamental structure is the roughness, which is the nature of the interfaces. For the interfaces with large roughness, one has to solve the NS equation by using mathematical tools such as the finite element method. And for the interfaces with small roughness, the rough interfaces can be treated as flat ones with the roughness involved in the slip length of the flat interfaces. So, for the latter, a question arises. What is the slip length $b$ of the interfaces with the roughness involved? In this study, we will focus on the interfaces with small roughness and address the problem to get the $b$ relating to the roughness in a quantitative way.\\

The slip length $b$ is defined as the ratio of the shear viscosity $\eta$ of liquid to the friction coefficient $\kappa$ of the solid surface by $b=\eta /\kappa$. It is known that the $\eta$ is position dependent in the liquid~\cite{Bocquet3,Hu,Israelachvili,Chan,Georges}. Especially, the viscosity of liquid in the region close to the interfaces may be enhanced compared to the viscosity in the region away from the interfaces. The discrepancy of the viscosities in the two regions leads to the shift of the exact  location of the BC applied away from the solid surfaces by a few molecular layers~\cite{Bocquet3,Chen}. However, in this study, we consider that the fluid is Newtonian and the $\eta$ is  constant in the whole computational domain for simplicity, since we only focus on the effect of the roughness of the interfaces on the slip length $b$. \\

The $\kappa$ is used to show the resistance of the solid surfaces to the liquid. Generally, two mechanisms are contributed to the $\kappa$. One mechanism is of the chemical interaction between the liquid molecules and the solid molecules at the solid-liquid interfaces, like that the attractive or repulsive force to aqueous liquids depends on whether the solid surfaces are hydrophobic or hydrophilic  respectively. The other mechanism is due to the fine structures of the interfaces, such as the roughness of the interfaces. The collusions between the liquid molecules and the rough interfaces bring the friction force to the liquid. We denote the friction coefficient contributed from the chemical interaction by $\kappa_w$ and the contribution of roughness of the interfaces by $\kappa_r$ in the following. For an ideally flat interface, the chemical interaction contributes totally to the $\kappa$ with $\kappa=\kappa_w$ and $\kappa_r=0$. However, for a real interface, the roughness is inevitable and $\kappa_r$ is nonzero. In such a real case, the friction forces respecting the two mechanisms then are obtained by $f_{w}=\kappa_{w}v_s$ and $f_{r}=\kappa_{r}v_s$ with $v_s$ the fluid velocity at the interfaces in the moving direction. Due to the additivity of the friction forces, we get the total friction coefficient as $\kappa=\kappa_w+\kappa_r$. \\

The determination of friction force of the solid surfaces, hence the $b$, for the confined liquid has been a long-standing question and has been heavily studied by many authors~\cite{Cottin1,Thompson1,Thompson2,Chan, Georges,Thompson,Smith,Walz,Liu, Ou1,Ou2,Choi1,Choi2,Callies,Truesdell,Bocquet4,Joseph,Lauga,Richardson,Cottin4}. Experimental and theoretical results, as well as the Molecular Dynamics Simulations, have been reported. In those studies, the roughness of the solid surfaces causing the no-slip BC has been pointed out and later confirmed by the Molecular Dynamics Simulations ~\cite{Richardson,Cottin1,Cottin4}. However, the role of the interface roughness in the slip BC actually is still lack. A very interesting result basing on the linear response theory has been obtained, in which the structure factor of the liquid in the region close to the interfaces is involved in the $\kappa$~\cite{Bocquet3,Bocquet2}. However, the structure factor does not have an explicit relation to the roughness of the interfaces in that study. In order to complete the understanding of the flow boundary conditions for confined liquid, it is necessary to bridge the slip length $b$ to the roughness of the solid surfaces, which is our goal in this study.\\   

\section{theory}
We consider a model that simple liquid is confined by two parallel solid plates. We set the coordinate with the $z$ axis normal to the plates and the $x$ axis parallel to the moving direction of the liquid. The two plates both are infinitely long to get rid off the ending effects of the structure. The two solid-liquid interfaces are located at $z=+h$ and $z=-h$ for the upper and lower plates respectively. The friction force applied on the fluid is considered to be opposite to the moving direction of the $x$ axis, while the component of the friction force along the $y$ direction is averaged to be zero. For simplicity, we think the interfaces are flat along the $y$ direction, and then the roughness of the interfaces is the function of $x$ only. We use the function $\zeta_u(x)$ to describe the roughness of the upper interface and use $\zeta_d(x)$ for the lower one. The two roughness functions both are averaged to be zero over the interfaces at their locations of $z=+h$ and $z=-h$ respectively.
 
\subsection{Fluid Velocity}
The complete Navier-Stokes(NS) equation for the fluid in the channel has the form of
\begin{equation}
\label{NSequ}
\eta \nabla \times \nabla \times \vec{v}+\rho \frac{d \vec{v}}{d t}=-\nabla p+\rho \vec{g} -\rho \vec{v} \cdot \nabla \vec{v}+\frac{4 \eta}{3}\nabla (\nabla \cdot \vec{v}).
\end{equation} 
Here, $\vec{v}$ is the fluid velocity and $\rho$ the liquid density. The $\rho$ and the $\eta$ have been assumed to be uniform for the liquid in the channel, since we focus on the effects from the solid surfaces only. In the right hand side of the eq.(\ref{NSequ}), the first two terms represent the force of the gradient of pressure and the gravitational force respectively. The third term is for the convect and the final term is due to the compressibility of the fluid. In the eq.( \ref{NSequ}), the equality of $\nabla^2 \vec{v}=\nabla(\nabla \cdot \vec{v})-\nabla \times \nabla \times \vec{v}$ has been applied. We think the roughness of the interfaces is not too large to bring the convect effect in the channel and the fluid is incompressible. Thus, we drop off the last two terms on the right hand side of eq.( \ref{NSequ}). And for simplicity, we use $\vec{f}$ to represent the two force terms by $\vec{f}=-\nabla p+\rho \vec{g}$. Based on the above considerations, the complete NS equation is reduced to be
\begin{equation}
\label{NSequ2}
\nabla \times \nabla \times \vec{v}+\frac{\rho}{\eta} \frac{d \vec{v}}{d t}=\frac{1}{\eta}\vec{f}.
\end{equation} 
Now we apply the time Fourier transformations of $\vec{U}(\omega)=\int _{-\infty}^{+\infty} \vec{v} e^{i\omega t}dt$ and $\vec{F}(\omega)=\int _{-\infty}^{+\infty} \vec{f} e^{i\omega t}dt$ on the both sides of the eq.(\ref{NSequ2}) and transform the equation in the frequency domain to be
\begin{equation}
\label{NSU}
\nabla \times \nabla \times \vec{U}-i \frac{\rho}{\eta}\omega \vec{U}=\frac{1}{\eta} \vec{F}.
\end{equation}
We introduce a dyadic Green function $\overset{\leftrightarrow}{G}(\vec{r}, \vec{r}',\omega)$, satisfying the following equation
\begin{equation}
\label{Green}
\nabla \times \nabla \times \overset{\leftrightarrow}{G}-i \frac{\rho}{\eta}\omega \overset{\leftrightarrow}{G}=\overset{\leftrightarrow}{I}\delta(\vec{r}-\vec{r}'),
\end{equation}
with $\overset{\leftrightarrow}{I}$ the dyadic unit and $\vec{r}(\vec{r}')$ the position vector in the channel. We consider the equations of one fixed frequency and drop off the notation of $\omega$ for convenience. Taking the dot product of $ \overset{\leftrightarrow}{G}$ on both sides of the eq.(\ref{NSU}), and taking the dot product of $\vec{U}$ on both sides of eq.(\ref{Green}), finally subtracting the two equations after the dot products, we obtain 
\begin{equation}
\label{subtraction}
[\nabla \times \nabla \times \vec{U}]\cdot \overset{\leftrightarrow}{G}-\vec{U}\cdot [\nabla \times \nabla \times \overset{\leftrightarrow}{G}]=\frac{1}{\eta} \vec{F}\cdot \overset{\leftrightarrow}{G}-\vec{U}\cdot \overset{\leftrightarrow}{I}\delta(\vec{r}-\vec{r}').
\end{equation}
Integrating both sides of the eq.(\ref{subtraction}) over the computational domain, we get the solution of $\vec{U}$ for the liquid with the form of
\begin{align}
\label{solutionU}
\vec{U}(r')=&-\int \hspace{-0.3cm} \int [(\hat{n} \times \vec{U}(r))\cdot (\nabla \times \overset{\leftrightarrow}{G}(r,r'))\nonumber \\
&+ (\hat{n} \times \nabla \times \vec{U}(r))\cdot \overset{\leftrightarrow}{G}(r,r')] ds \nonumber\\
&+\int \hspace{-0.3cm} \int \hspace{-0.3cm} \int \frac{1}{\eta}\vec{F}(r)\cdot \overset{\leftrightarrow}{G}(r,r')dV .
\end{align}
On the right hand side of the eq.(\ref{solutionU}), the first integral of $\iint ds$ is the integral over the total surface of the computational domain, which is obtained by the applying of the Green theorem on the volume integral of the left hand side of the eq.(\ref{subtraction}). The Green theorem has been shown in the Appendix A. The $\hat{n}$ is the unit vector outward normal to the interfaces. Note that the direction of the $\hat{n}$ varies on the interfaces due to the roughness.\\

For simplicity, we assume the inlet of the channel along the $x$ direction is periodic to the outlet. Thus the integral over the inlet and outlet cross-section surfaces would be cancelled. Similar cancellation is hold for the integral over the cross-section surfaces normal to the $y$ direction due to the uniformity along the direction. In this way, the term of surface integral in the  eq.(\ref{solutionU}) remains as the integral over the solid-liquid interfaces only. Without losing the generality, we will focus on only the upper interface to figure out the contribution of the roughness to the slip length, since the similar treatment can be applied to the lower interface. For convenience, we use $K$ to represent the integrand of the surface integral by
\begin{align}
\label{K}
K=- (\hat{n} \times \nabla \times \vec{U})\cdot \overset{\leftrightarrow}{G}- (\hat{n} \times \vec{U})\cdot (\nabla \times \overset{\leftrightarrow}{G}).
\end{align}

\subsection{Boundary Condition}
The essential of the NBC is that the friction force applied on the fluid per unit length should be balanced by the shear stress of the fluid at the interface. The friction force and the shear stress have the same magnitude but the opposite directions. We denote the velocity component of the fluid parallel to the interfaces by $v_{||}$. Then in the case that the solid-liquid interfaces are ideally flat, the shear stress is along the $x$ direction and has been well defined as the product of the liquid viscosity $\eta$ and the normal gradient of the $v_{||}$ to the interface, while the friction force is along the $-x$ direction and is the product of the friction coefficient $\kappa$ and the $v_{||}$ at the interface. Thus, for the ideal interfaces, the force balance can be expressed by $\eta \frac{\partial v_{||}}{\partial z}=-\kappa_w v_{||}$ valid at the interfaces, with the minus meaning the friction force resists the fluid flow. This is exactly the NBC. Here, only the $\kappa_w$ appears in the NBC, since no any effect of roughness of the interfaces contributes to the friction coefficient in such ideal case. The slip length then is defined as $b_w=\eta/\kappa_w$.\\

In the non-ideal case that the solid-liquid interfaces are not ideally flat but with roughness, the NBC can be expressed locally since the force balance is still hold locally on the interfaces. We will average the fluid velocity eq.(\ref{solutionU}) over the rough interfaces to get an effective fluid velocity. From an alternative view, the effective fluid velocity can be solved from the NS equation by imposing an effective NBC(ENBC) on ideally flat interfaces. The ENBC has absorbed the effect of roughness of the non-ideal interfaces into the friction coefficient $\kappa$ of the ideally flat interfaces. The total friction coefficient $\kappa$ of the flat interfaces after the average then is the sum of $\kappa_w$ and $\kappa_r$ as mentioned in the introduction. The total slip length $b$ of the flat interfaces after the average can be written as $b=b_wb_r/(b_w+b_r)$ with $b_w=\eta/\kappa_w$ and $b_r=\eta/\kappa_r$ by using the equalities of $\kappa=\kappa_w+\kappa_r$ and $b=\eta/\kappa$. It can be seen that for flat interfaces with $b_r \rightarrow \infty$, the contribution of chemical interaction dominates the friction force with $b=b_w$. On the other side, when $b_w \rightarrow \infty$, the friction force then is dominated by the roughness of the interfaces with $b=b_r$. So, after getting the total slip length $b$ of the confined liquid, the rough solid-liquid interfaces can be replaced by flat ones in an effective way, which simplifies the problems of the fluid dynamics in real systems with the roughness having been involved in the $b$. In the following study, we will set the $b_w$ as a priori since the $b_w$ depends on the natural interaction between the solid surfaces and fluid molecules, and focus on the informations of the roughness.\\

We start from the upper rough interface. For convenience, we denote $\hat{\tau}$ the unit vector on the interface along the tangential direction in the $x,z$ plane and  perpendicular to the $\hat{n}$. The $\hat{\tau}$ can be expressed by $\hat{\tau}=\frac{\hat{n} \times \vec{U} \times \hat{n}}{|\vec{U} \times \hat{n}|}$. The local force balance on the rough interfaces, hence the NBC, can be found generally to be
\begin{align}
\label{NC1}
& \vec{U}\cdot \hat{n}=0,\\
\label{NC2}
& \eta \hat{n}\cdot (\nabla \vec{U}) \cdot \hat{\tau}=- \kappa_w \vec{U} \cdot \hat{\tau},\\
\label{NC3}
& \hat {n} \cdot (\nabla \vec{U}) \cdot (\hat{\tau} \times \hat{n})=0.
\end{align} 
In the general NBC, eq.(\ref{NC1}) means the component of $\vec{U}$ normal to the interfaces vanishes, showing that the liquid can not penetrate the solid surface. In the eq.(\ref{NC2}), the term on the left hand side is the shear stress which is  along the $\hat{\tau}$ direction and is on the plan normal to $\hat{n}$. The term on the right hand side of the eq.(\ref{NC2}) is the friction force. The final condition eq.(\ref{NC3}) means no stress along the $y$ direction according to our model.\\

It is emphasized here that the slip length used in the eq.(\ref{NC2}) is the $b_w=\eta/\kappa_w$ and no $\kappa_r$ is involved. This is because the force balance considered now is hold locally. The involvement of $\kappa_r$ in the slip length $b$ will be obtained after the averaging of the eq.(\ref{solutionU}) over the rough interfaces by using the general NBC. For the application of the general NBC, we need to modify the eq.(\ref{NC1}, \ref{NC2}, \ref{NC3}) to have the following forms
\begin{align}
\label{NC11}
& \vec{U} \cdot \hat{n}=0,\\
\label{NC22}
&  \vec{U} \cdot [(\hat{n}\cdot \nabla)\vec{U}]=-\frac{1}{b_w} \vec{U} \cdot \vec{U},\\
\label{NC33}
& [(\hat{n}\cdot \nabla)\vec{U}]\cdot [\hat{n}\times \vec{U}]=0.
\end{align}
The above modification can be found in the Appendix B.

\subsection{Resolution of $K$}
We need to resolute the $K$ defined in eq.(\ref{K}) into several terms to figure out the effect of the roughness. Before the resolution, we decompose the vector $\nabla \times \vec{U}$ into two vectors with one vector  along the $y$ axis and the other vector in the $x, z$ plan, having
\begin{align}
\label{nablaU}
\nabla \times \vec{U}=&\frac{[(\nabla \times \vec{U})\cdot (\hat{n} \times \vec{U})](\hat{n} \times \vec{U})}{(\hat{n}\times \vec{U})\cdot (\hat{n}\times \vec{U})}\nonumber\\
+& \frac{(\hat{n}\times \vec{U})\times (\nabla \times \vec{U})\times (\hat{n}\times \vec{U})}{(\hat{n}\times \vec{U})\cdot (\hat{n}\times \vec{U})}.
\end{align}   
By using the eq.(\ref{NC11}, \ref{NC22}, \ref{NC33},\ref{nablaU}), the vector $n\times \nabla \times U$ in the $K$ can be split into three terms, reading
\begin{align}
\hat{n}\times \nabla \times \vec{U}=\vec{T}_1+\vec{T}_2+\vec{T}_3,
\end{align}
with
\begin{align}
&\vec{T}_1=\frac{- \vec{U}\cdot \vec{U}(\hat{n} \times \hat{n} \times \vec{U})}{b_w( \hat{n}\times \vec{U})\cdot ( \hat{n}\times \vec{U})},\nonumber\\
&\vec{T}_2=\frac{-\hat{n}\cdot ((\vec{U}\cdot \nabla)\vec{U})(\hat{n} \times \hat{n} \times \vec{U})}{( \hat{n}\times \vec{U})\cdot ( \hat{n}\times \vec{U})},\nonumber\\
&\vec{T}_3=\frac{(\hat{n} \times \vec{U}) [\hat{n}\cdot ((\hat{n}\times \vec{U})\cdot \nabla)\vec{U}]}{(\hat{n}\times \vec{U})\cdot (\hat{n}\times \vec{U})}.\nonumber 
\end{align}
The details for the above derivation can be found in the Appendix C. We write the $\vec{U}$ in the form of $\vec{U}=M\hat{\tau}$ with $M$ the magnitude of $\vec{U}$. Then, we can further simplify the terms of $\vec{T}_1, \vec{T}_2, \vec{T}_3$ to be
\begin{align}
\label{T1}
&\vec{T}_1=\frac{M\hat{\tau}}{b_w},\\
\label{T2}
&\vec{T}_2=M\hat{\tau}(\hat{n}\cdot (\hat{\tau}\cdot \nabla)\hat{\tau}), \\
\label{T3}
&\vec{T}_3=M(\hat{n}\times \hat{\tau})(\hat{n}\cdot ((\hat{n}\times \hat{\tau})\cdot \nabla)\hat{\tau}).
\end{align}
In the above simplification, the equality of $\hat{n}\cdot \hat{\tau}=0$ has been applied. The similar treatment can be applied to the vector $\hat{n}\times \vec{U}$ in the $K$, leading to
\begin{align}
\hat{n}\times \vec{U}=\vec{T}_4=M\hat{n}\times \hat{\tau}.
\end{align}
Thus, the integrand $K$ defined in the eq.(\ref{K}) can be written as
\begin{align}
K=- (\vec{T}_1+\vec{T}_2+\vec{T}_3)\cdot \overset{\leftrightarrow}{G}- \vec{T}_4\cdot (\nabla \times \overset{\leftrightarrow}{G}).
\end{align}
It can be understood that for an ideally flat interface where the $\hat{\tau}$ is not a variable but a constant vector along the $x$ direction on the interface, the two terms of $\vec{T}_2$ and $\vec{T}_3$ both disappear because the operator $\nabla$ applied on the constant vector $\hat{\tau}$ causes zero in the eq.(\ref{T2}, \ref{T3}). And then only the terms of $\vec{T}_1\cdot \overset{\leftrightarrow}{G}$ and $\vec{T}_4\cdot (\nabla \times \overset{\leftrightarrow}{G})$ remain in the $K$ for the ideally flat interface. Therefor, the terms of $(\vec{T}_2+\vec{T}_3)\cdot \overset{\leftrightarrow}{G}$ in the $K$ are originated from the roughness of the interface.

\subsection{Perturbation Expansion}
We have used the function $\zeta_u(x)$ to describe the roughness of the upper interface. For convenience, we omit the subscript to denote the roughness by $\zeta(x)$. The local unit vectors of the interface can be expressed in the term of $\zeta(x)$, reading
\begin{align}
\label{geometr1}
&\hat{n}=\frac{1}{N}(-\frac{\partial \zeta}{\partial x},~~0, ~~1),\nonumber\\
&\hat{\tau}=\frac{1}{N}(1,~~0,~~ \frac{\partial \zeta}{\partial x}),
\end{align}
with
\begin{align}
\label{geometr2}
N=\sqrt{(\frac{\partial \zeta}{\partial x})^2+1}.
\end{align}
Now we need to average the fluid velocity $\vec{U}$ over the rough interface to get an effective velocity $\vec{W}=<\vec{U}>$. Alternatively, the fluid vector $\vec{W}$ can be solved from the NS equation by imposing the ENBC on a flat interface. The ENBC for such flat interface should have the form of 
\begin{align}
\label{NCi}
W_z=W_y=0,~~\frac{\partial W_x}{\partial z}=-\frac{W_x}{b}.
\end{align}
Here, the total slip length $b$ has been used to comprise the two effects of the interface, the chemical interaction and the roughness. The dyadic Green function for the effective system with the flat interface is denoted by $\overset{\leftrightarrow}{g}$. The boundary conditions for the $\overset{\leftrightarrow}{g}$ are the same as the eq.(\ref{NCi}) only by replacing the $W_i$ with the $g_{ij}$ where $i$ and $j$ represent the $x$, $y$ and $z$ coordinates. Note that $g_{ij}$ is the component of $\overset{\leftrightarrow}{g}$, meaning that the influence of the source of $i$ component is on the object of $j$ component. We expand the $\vec{U}$ and $\overset{\leftrightarrow}{G}$ of the rough interface in the terms of $\vec{W}$ and $\overset{\leftrightarrow}{g}$ of the flat interface respectively. Since the friction force considered is parallel to the $x$ direction and the $y$ component of the friction force is negligible, we only consider the $x$ component of $\vec{U}$ for the expansion and neglect the $y$ and $z$ components. The latter two components are averaged to be zero over the rough interface. In this way, for the $\overset{\leftrightarrow}{G}$, we only consider the components of $G_{xi}$. We have noted that the roughness of the interface should not be too large, or the roughness will bring the convection and turbulence effects to the system. Under the condition of $|\zeta/b|<<1$, the perturbation expansion of the $\vec{U}\cdot \hat{x}$ and the $G_{xi}$ read
\begin{align}
\label{expan}
&M\hat{\tau}\cdot \hat{x}=W_x+\zeta\frac{\partial W_x}{\partial z}=W_x(1-\frac{\zeta}{b}),\nonumber \\
&G_{xi}=g_{xi}+\zeta\frac{\partial g_{xi}}{\partial z}=g_{xi}(1-\frac{\zeta}{b}).
\end{align}
In above derivation, the ENBC of eq.(\ref{NCi}) have been applied.\\

In the following, we expand the $K$ in the terms of $W$ and $g$. For the first term $\vec{T}_1\cdot \overset{\leftrightarrow}{G}$ of the $K$, we get
\begin{align}
\label{T1expan}
\vec{T}_1\cdot G_{xi}\hat{x}\hat{i}=\frac{W_x}{b_w}(1-\frac{\zeta}{b})^2g_{xi}\hat{i}.
\end{align}
By substituting the eq.(\ref{geometr1},\ref{expan}) into the second term of the $K$, we get
\begin{align}
\label{T2expan}
\vec{T}_2\cdot G_{xi}\hat{x}\hat{i}&=W_x(1-\frac{\zeta}{b})^2(\hat{n}\cdot (\hat{\tau}\cdot \nabla)\hat{\tau}) g_{xi}\hat{i}\nonumber \\
&=W_x(1-\frac{\zeta}{b})^2\frac{\partial ^2 \zeta}{\partial x^2} N^{-3}g_{xi}\hat{i}
\end{align}
We observe that the vectors $T_3$ and $T_4$ both are along the $y$ direction, which have negligible contribution to the effect of the roughness as mentioned in this work. Thus, we will figure out the roughness information only from the two terms of eq.(\ref{T1expan}) and eq.(\ref{T2expan}).

\subsection{Slip Length}
The surface integral of the eq.(\ref{T1expan}) and eq.(\ref{T2expan}) obtained from the eq.(\ref{solutionU}) can be effectively equivalent to
\begin{align}
<\int\int[\vec{T}_1\cdot G_{xi}\hat{x}\hat{i}+\vec{T}_2\cdot G_{xi}\hat{x}\hat{i}]ds>=\int \int\frac{W_x}{b}g_{xi}\hat{i}dxdy.
\end{align}
The right hand side of the above equation is the surface integral over the flat interface after the average and can be understood from the eq.(\ref{T1}) with the eq.(\ref{T2}) and eq.(\ref{T3}) vanishing at the flat interface. Here, $b$ is used to comprise the both effects of the chemical interaction and the roughness for the flat interface. Therefore, we get the equation for the total slip length $b$, reading
\begin{align}
\label{bequ}
<\frac{N}{b_w}(1-\frac{\zeta}{b})^2+\frac{\partial ^2 \zeta}{\partial x^2}N^{-2}(1-\frac{\zeta}{b})^2>=\frac{1}{b}
\end{align}
by using the $\int \int ds=\int\int N dx dy$. The factor of $W_xg_{xi}\hat{i}dxdy$ on both sides of the equation are canceled, meaning that the $b$ is a character of the confined liquid, and is independent on the velocity of fluid and time. For simplicity, we denote $A=\frac{1}{N^2}\frac{\partial ^2 \zeta}{\partial x^2}+\frac{N}{b_w}$. After averaging both sides of the eq.(\ref{bequ}) statistically over the interface, we solve the slip length $b$ as
\begin{align}
\label{resultb}
b=\frac{2<\zeta A>+1+\sqrt{(2<\zeta A>+1)^2-4<A><\zeta^2A>}}{2<A>}.
\end{align}
This is the main result of this work.

%\nocite{*}
\section{Results and Discussions}
The topography of the rough interface is described by the $\zeta(x)$ together with the correlation function of the roughness. For simplicity, we set that the correlation function is Gaussian, which reads $<\zeta(x')\zeta(x'+x)>=\sigma^2 e^{-x^2/l^2}$. Here, $\sigma$ is the  variance of the roughness. The $x$ is the distance between two positions on the interface for the correlation. And the $l$ is the correlation length, meaning that the two positions with their distance larger than $l$ lose their correlation. The interface with a longer $l$ has a smoother topography. In the expression eq.(\ref{resultb}) of $b$, the information of $l$ has been reflected by the factor of $\frac{\partial ^2 \zeta}{\partial x^2}$ in the $A$. We have used Mont Carlo method to simulate various rough interfaces by varying $\sigma$ and $l$. The lattice parameter of the solid surface is used for the dimensionless. The physical quantities in this work are evaluated statistically over the rough interfaces. Fig. 1 shows that the $<N>$ is functional of the correlation length $l$ of the interface by varying the roughness variance $\sigma$. 
\begin{figure}[!hbp]
\centering
\includegraphics[width=0.5\textwidth]{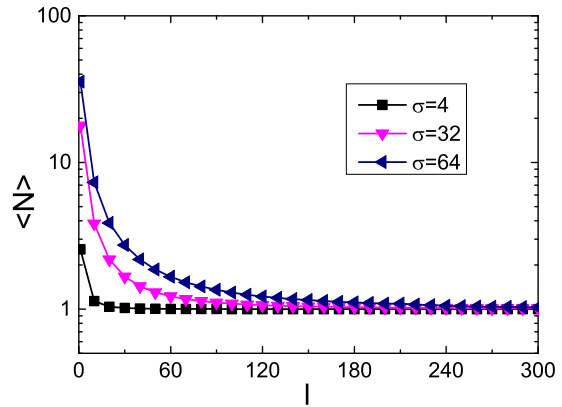}
\caption{Statistical average of $N$ over interfaces, noted by $<N>$, as a function of the correlation length $l$ with various roughness variances $\sigma$. The $N$ has been defined in the text by eq.(\ref{geometr2}). And the $<N>$ is plotted in the logarithmic scale. }
\end{figure}
In the figure, we find that the $<N>$ deviates from the number of one in the range of $l$ close to zero. Larger the $\sigma$, stronger the deviation of the $<N>$. With the increasing of $l$, the $<N>$ decreases and approaches to the unit, indicating that the interface is becoming smooth with $\frac{\partial \zeta}{\partial x}$ going to zero and $\hat{\tau}$ deviating from the unit vector $\hat{x}$ slightly. A rougher interface with a larger $\sigma$ need a larger $l$ to smooth the interface and to recover the $<N>$ to the unit. In the fig.1, the deviation of the $<N>$ actually means that the $l$ and the $\sigma$ both play their roles in determining the slip length. And the $l$ gradually weakens its effect in the determination with the increasing of $l$. Finally, only the $\sigma$ remains to determine the slip length if the $l$ goes to the limit of infinity.\\

\begin{figure}[!hbp]
\centering
\includegraphics[width=0.4\textwidth]{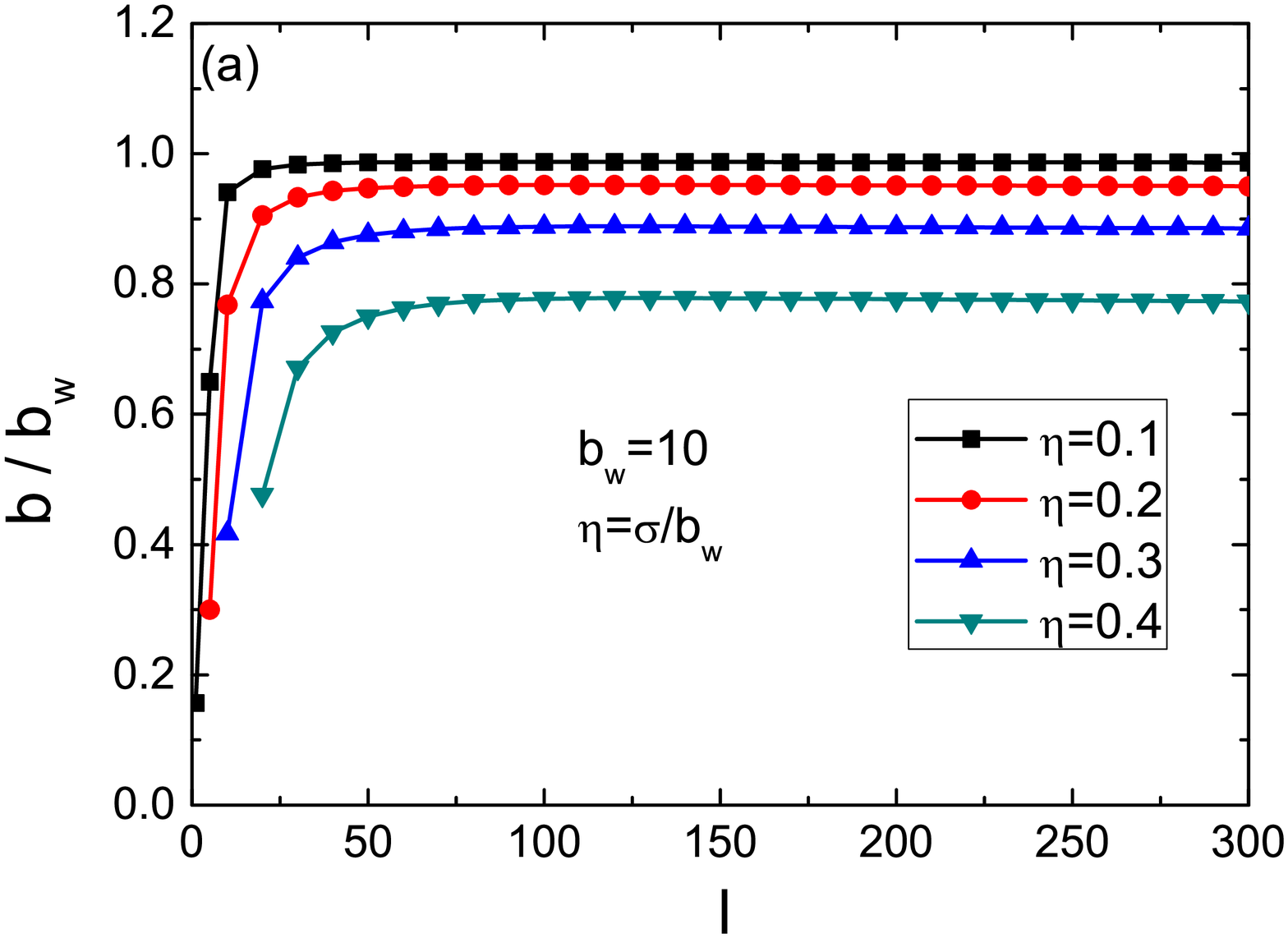}
\includegraphics[width=0.4\textwidth]{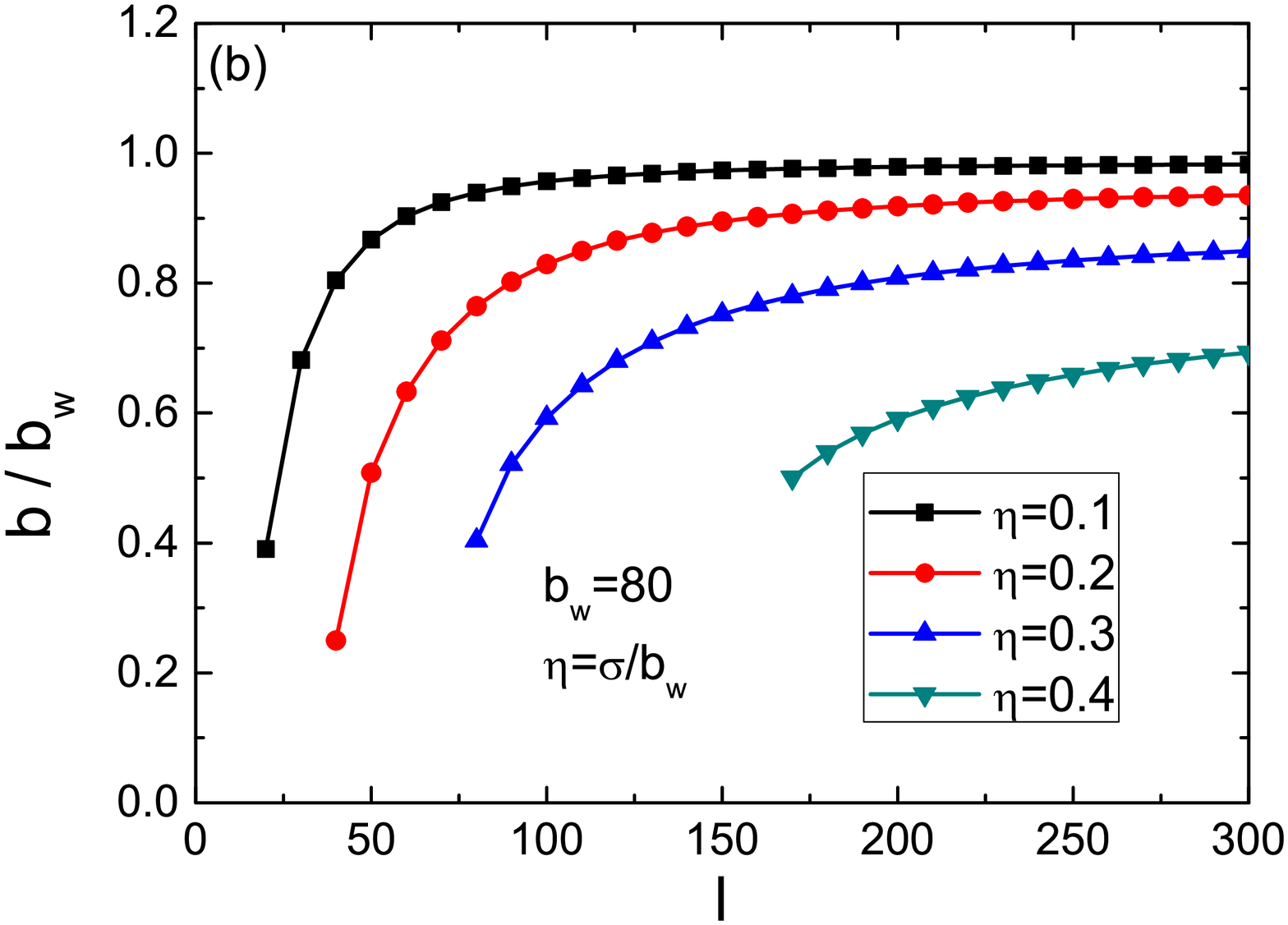}\\
\includegraphics[width=0.4\textwidth]{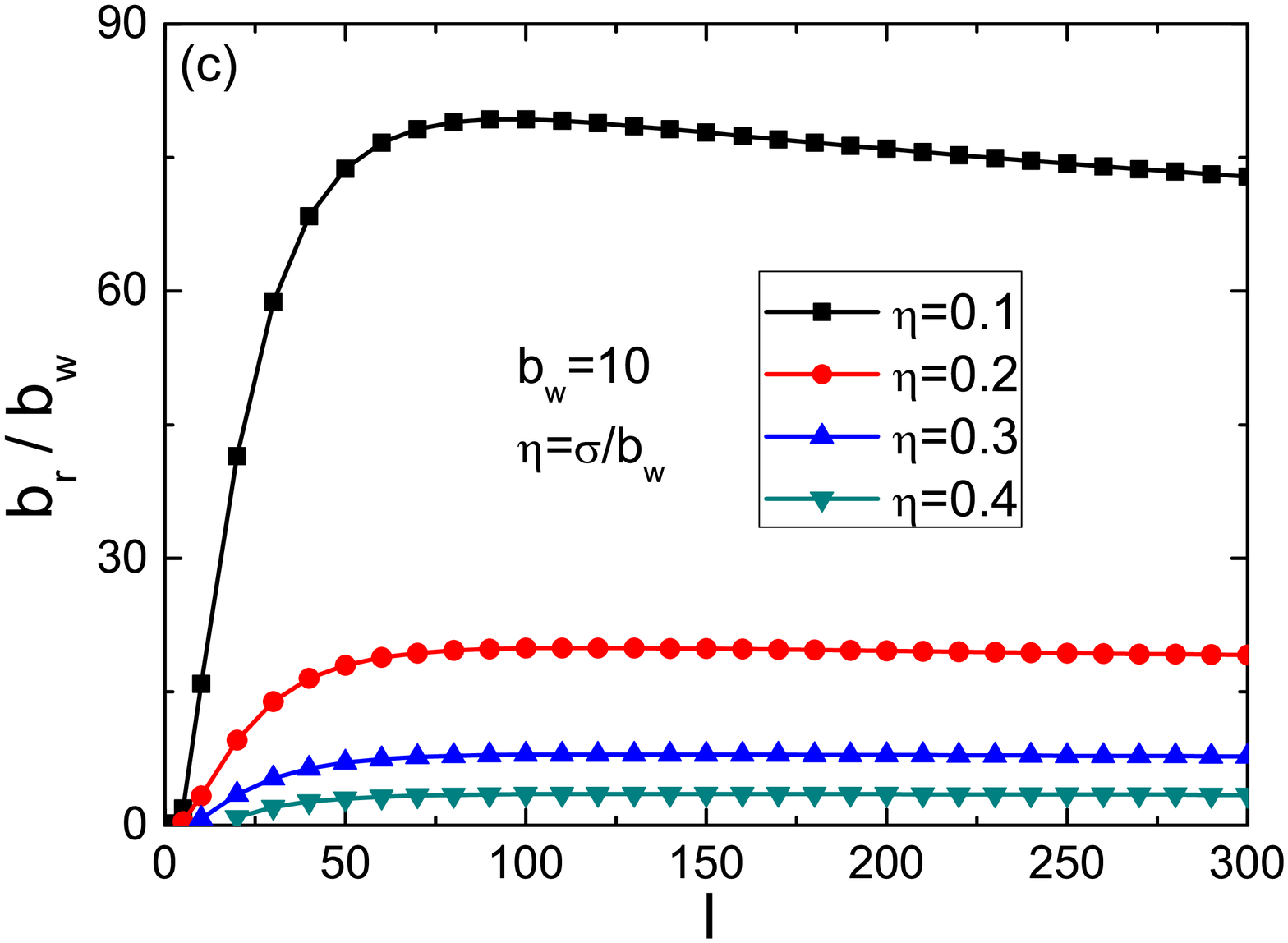}
\includegraphics[width=0.4\textwidth]{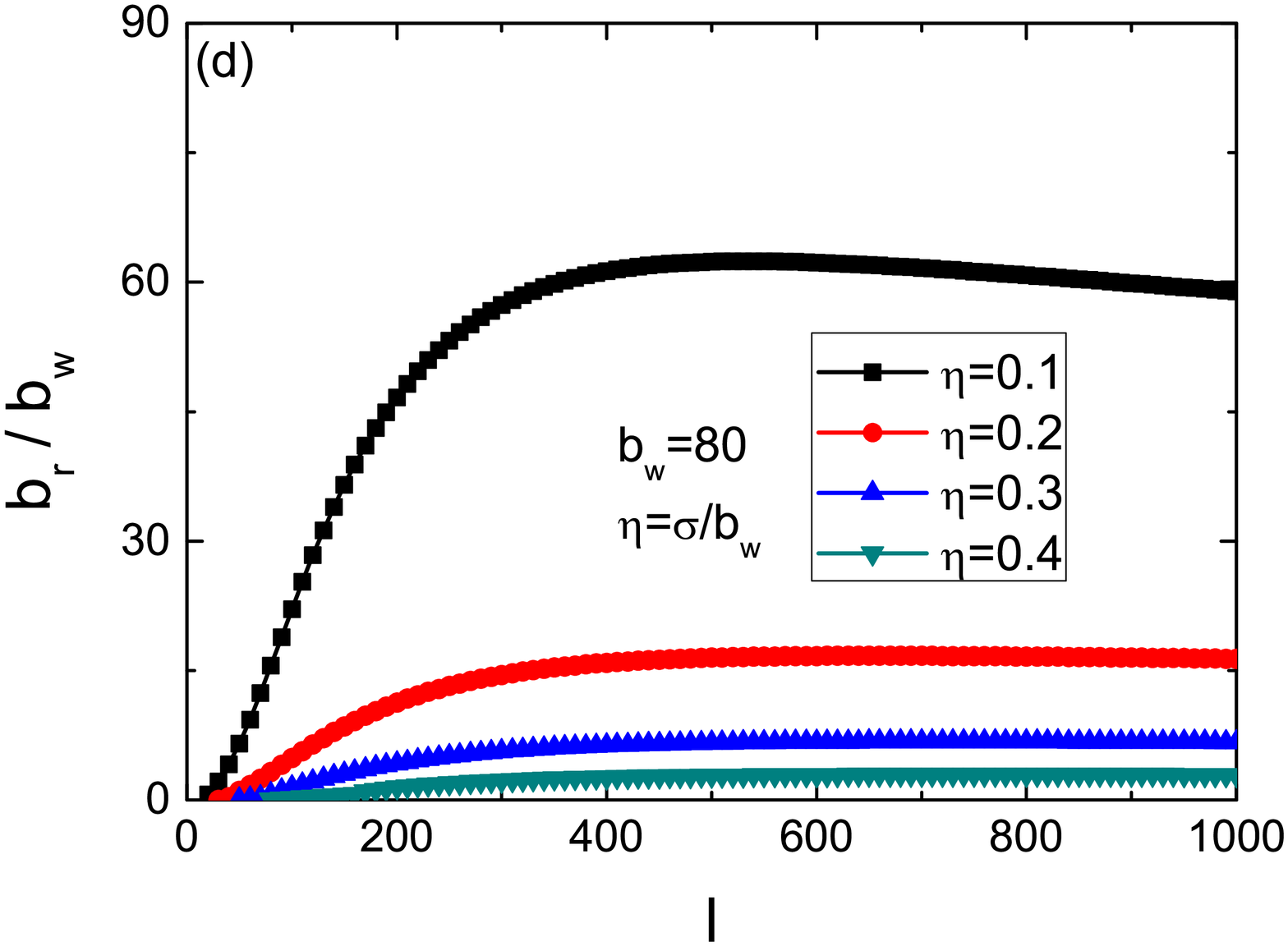}
\caption{Slip lengths as functions of the correlation length $l$ with various roughness variances $\sigma$. (a) and (b) are for the total slip length $b$ of the system while the (c) and (d) are for the slip length $b_r$ contributed from the roughness of interfaces. In the plot, the slip lengths and $\sigma$ are all renormalized by the slip length $b_w$ contributed from the chemical interaction of the interfaces.}
\end{figure}
Fig. 2 shows the slip lengths as functional of $l$ with various $\sigma$. It should be noted that in the eq.(\ref{expan}) the $|\zeta/b| <<1$ is used for the perturbation expansion instead of $|\zeta|<<1$. Thus, we use $\sigma/b_w$ to represent the variance of the roughness in the plot instead of $\sigma$ itself, considering that the $b$ is equivalent to $b_w$ in the extreme case of $\sigma=0$. We apply $\sigma/b_w \le 0.4$ in the plot to verify the perturbation expansion.
Fig.2(a) and (b) are for the total slip length $b$ basing on the eq.(\ref{resultb}), while the fig.2(c) and (d) are for the slip length $b_r$ contributed from the roughness only. The $b_r$ is obtained from $\frac{1}{b_r}=\frac{1}{b}-\frac{1}{b_w}$. The slip lengths are renormalized by the $b_w$ in the figures for clarity. The $b_w$ has been indicated in each figure. In the fig.2(a) and (b), it can be found that the $b$ increases dramatically with the increasing of the $l$ and then goes to a saturation value. The saturation value of the $b$ is independent on the $l$, but decreases with the increasing of $\sigma$ as expected. The increasing of $b$  with the $l$ emphasizes that in order to increase the $b$ it is essential not only by choosing proper material for the channel fabrication, but also by smoothing the solid surfaces in the fabrication. The $b_r$ in the fig.2(c) and (d) are calculated from the data of fig.2(a) and (b) respectively, showing the similar behaviors of $b$ in the fig.2(a) and (b). In the fig.2(d), we have appended the scale of $l$ to 1000 because the $b_r/b_w$ does not reach the saturation at the $l=300$. The fig. 2(c) and (d) show that the slip length contributed from the roughness is a function of $l$ and $\sigma$ in the range of $l$ close to zero, and finally is a function of $\sigma$ only when the $l$ goes to the limit of infinity. We need to note that in the fig.2 the scaling behaviors of the slip lengths are dependent on the ratio of $\sigma$ to $b_w$ rather than the $\sigma$ itself. And for the interfaces with large roughness where the perturbation expansion is invalid, say $\sigma/b_w>0.4$, rigorous numerical calculations such as the finite element method should be applied to solve the NS equation, which is out of the scope of this work. \\

\begin{figure}[!hbp]
\centering
\includegraphics[width=0.4\textwidth]{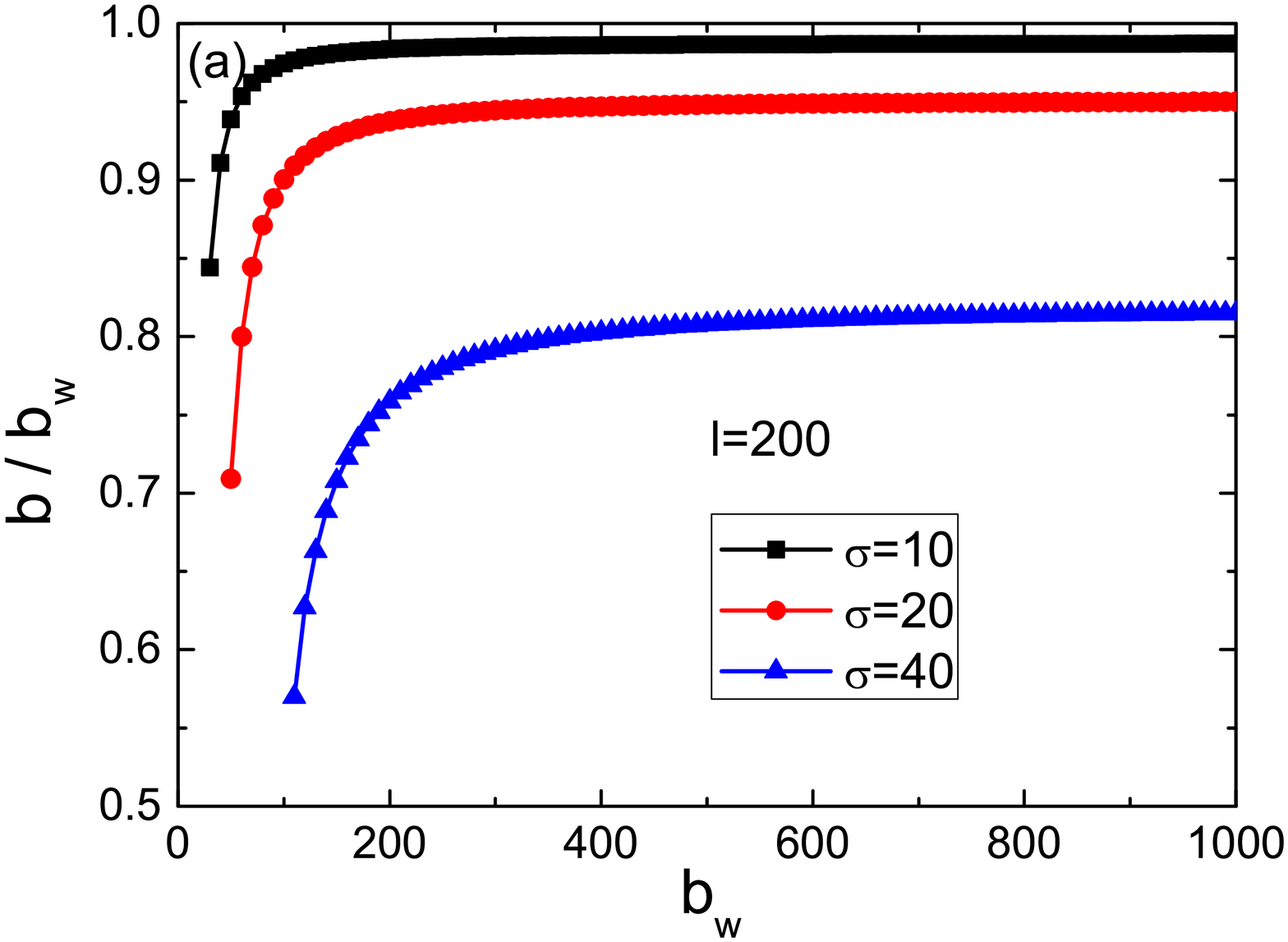}
\includegraphics[width=0.4\textwidth]{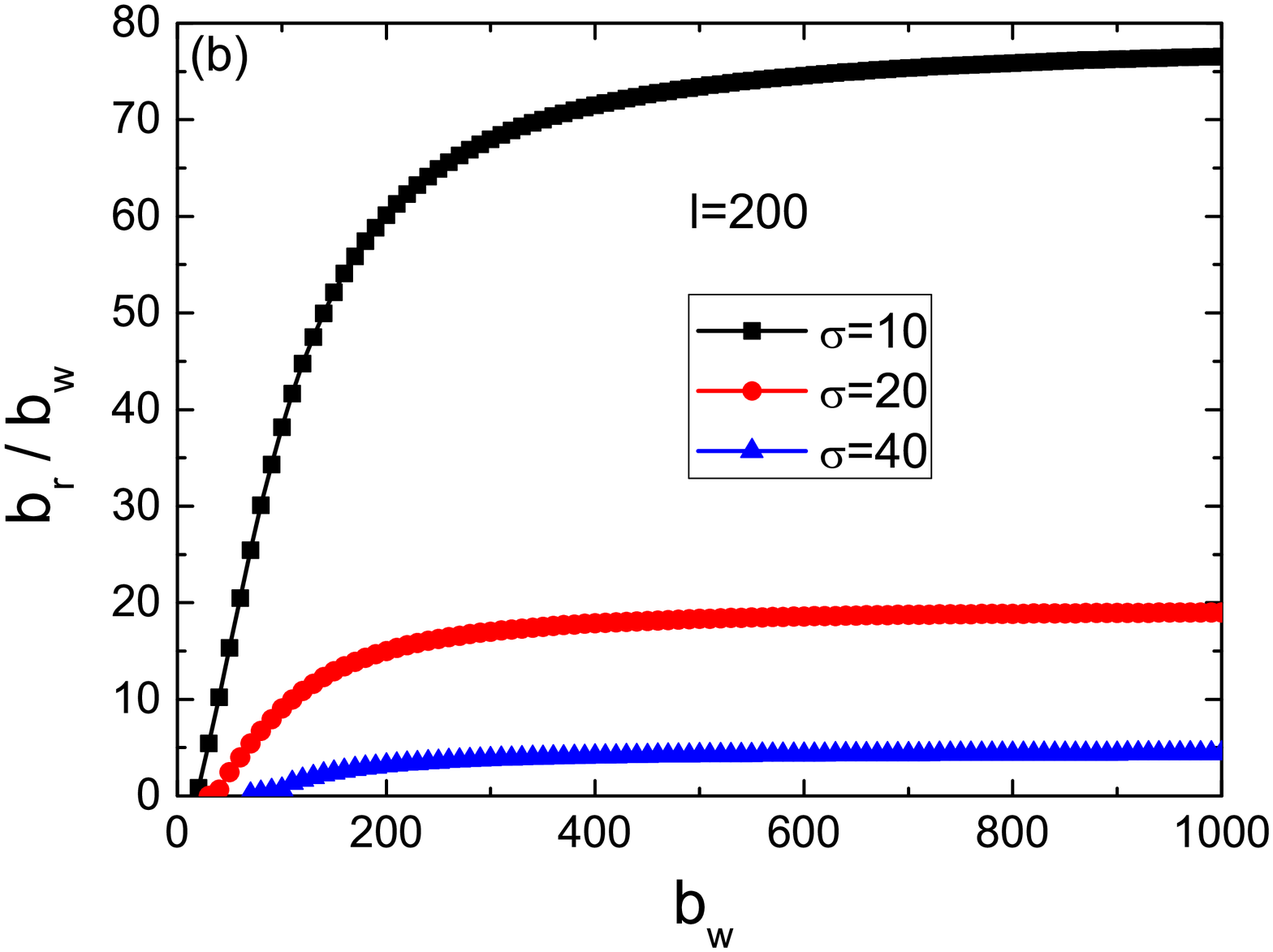}
\caption{(a)Proportion of $b$ to $b_w$. (b)Proportion of $b_r$ to $b_w$. In the plot, the $b$ and $b_r$ have been renormalized by the $b_w$ for clarity.}
\end{figure}
Before we figure out the relation between the the $b(b_r)$ and the $b_w$, we firstly make a simple analysis. We approximate the following equalities of $<A>=1/b_w$, $<\zeta A>=<\frac{\zeta}{N^2}\frac{\partial ^2 \zeta}{\partial x^2}>$ and $<\zeta^2 A>=<\zeta^2>/b_w$ and find that $b$ can be expressed by $b \sim b_w <f(\zeta, \frac{\partial ^2 \zeta}{\partial x^2})>$. That means $b$ is linearly proportional to $b_w$ and the roughness of the interface plays its role as the proportionality factor. Now we check the proportion in fig.3 with various $\sigma$. Here, the correlation length is fixed at $l=200$ in the plot and  $\sigma$ is used instead of the $\sigma/b_w$. Fig.3(a) shows that the ratio of the $b$ to the $b_w$ goes up to a constant with the increasing of $b_w$. The constant ratio confirms the linear proportion of the $b$ to the $b_w$ and the roughness plays as the proportionality factor. Larger the $\sigma$, smaller the ratio, meaning that the roughness weakens the effect of $b_w$. Fig.3(b) is for the $b_r$ calculated from the data of fig.3(a), also showing the linear behavior as the $b$. The relationship between the $b_r$ and the $b_w$ indicates that the roughness of the interface can not play the role itself in determining the $b$, but need to be with the $b_w$ together. That means if $b_w$ is infinite, $b_r$ must be infinite too and then is the $b$.\\

\section{conclusion}
In this work, we have studied the slip length of the confined liquid with small roughness of solid-liquid interfaces. The quantitative expression of the slip length has been obtained, in which the roughness of the interfaces and the chemical interaction between the liquid and the solid are involved. In the study, perturbation expansion of the interface roughness is required. The results we have obtained are threefold. First, the slip length $b$ of the confined liquid is linearly proportional to the slip length $b_w$ contributed from the chemical interaction. Second, the roughness of the interface plays its role as a proportionality factor for the $b$ and the $b_w$. And only with the chemical interaction together does the roughness have its effect on the $b$. Finally, large variance of the roughness decreases the $b$ and the increasing of the correlation length of the roughness increases the slip length in a dramatical way to a saturation value. Those results propose that we need to improve our techniques to smooth the solid surfaces by enlarging the correlation length of the roughness as well as decreasing the variance of the roughness for the channel fabrication besides to find proper materials with large $b_w$.\\

It has been revealed that some biological nanopores have peculiar features in the transport of water. Such as in the aquaporin channels as the key component of many biological processes, the permeability of water is larger than what is expected in classical nanofluidic framework by three orders of magnitude~\cite{Walz}. It has been suggested that the hydrophobic surface of the aquaporin channels is the potential mechanism to the large permeability of water in the channels. To supplement the suggestion, we think that the large correlation length of the channels may contribute to the water transport as have been studied in this work, even though how the positions are correlated in the aquaporin channels is not clear. What is more, the biological nanopores are active matters, in which the correlation between the two positions on the nanopore interfaces may behave in a complicated way rather than what does in the artificial nanofluidic frameworks. The consideration of the correlation length in the biological nanopores is the extension of this work and under the research.

\appendix
\section{}
The second kind Green theorem is 
\begin{align}
&\int \hspace{-0.3cm} \int \hspace{-0.3cm} \int [(\nabla \times \nabla \times \vec{U})\cdot \overset{\leftrightarrow}{G}-\vec{U} \cdot(\nabla \times \nabla \times \overset{\leftrightarrow}{G})]dV\nonumber\\
&=\int \hspace{-0.3cm} \int [(\hat{n} \times \vec{U})\cdot (\nabla \times \overset{\leftrightarrow}{G})+ (\hat{n} \times \nabla \times \vec{U})\cdot \overset{\leftrightarrow}{G}] ds.
\end{align}

\section{}
We need to modify the NBC eq.(\ref{NC1}, \ref{NC2}, \ref{NC3})for the application as the first step in splitting the $K$. The eq.(\ref{NC1}) can be written in the form of 
\begin{equation}
\label{MNC1}
 U_a n_a=0,
\end{equation}
with the Einstein's notation. Here, the subscript $a$ represents the coordinates $x,y,z$.\\

Then, we substitute the $\tau=\frac{\hat{n} \times \vec{U} \times \hat{n}}{|\vec{U} \times \hat{n}|}$ into the eq.(\ref{NC2}), and get rid of the common denominators on both sides of the equation. In this way, we get the modified  eq.(\ref{NC2}) reading
\begin{equation}
\label{MNC2}
\hat{n}\cdot (\nabla \vec{U}) \cdot [\hat{n} \times \vec{U} \times \hat{n}]=- \frac{1}{b_w} \vec{U} \cdot [\hat{n} \times \vec{U} \times \hat{n}].
\end{equation}
The left hand side of the eq.(\ref{MNC2}) can be simplified by using the Einstein's notation
\begin{align}
&\hat{n}\cdot (\nabla \vec{U}) \cdot [\hat{n} \times \vec{U} \times \hat{n}]\nonumber \\
&=(n_a\vec{i}_a)\cdot(\vec{i}_b \frac{\partial U_c}{\partial x_b}\vec{i}_c)\cdot(\epsilon_{efg}\vec{i}_e n_f \epsilon_{ghk}U_hn_k)\nonumber \\
&=\epsilon_{efg}\epsilon_{ghk}n_b \frac{\partial U_e}{\partial x_b}n_f U_hn_k \nonumber \\
&=(\delta_{eh}\delta_{fk}-\delta_{ek}\delta_{fh})n_b \frac{\partial U_e}{\partial x_b}n_f U_hn_k \nonumber \\
&=n_bU_e\frac{\partial U_e}{\partial x_b}=\vec{U} \cdot [(n\cdot \nabla)\vec{U}].
\end{align}
Here, the eq.(\ref{MNC1}) has been applied. Similarly, the right hand side of the eq.(\ref{MNC2}) can be simplified as $-\frac{1}{b_w} \vec{U} \cdot \vec{U}$. So the eq.(\ref{MNC2}) can be transformed as
\begin{equation}
\label{MNC21}
\vec{U} \cdot [(n\cdot \nabla)\vec{U}]=-\frac{1}{b_w} \vec{U} \cdot \vec{U}.
\end{equation}
This equation will be used for the splitting of the $K$.\\
 
The eq.(\ref{NC3}) then can be transformed to be
\begin{align}
&\hat {n} \cdot (\nabla \vec{U}) \cdot (\vec{U}\times \hat{n})=\epsilon_{bef} n_a n_e U_f \frac{\partial U_b}{\partial x_a}=0\nonumber \\
&\Rightarrow [(n\cdot \nabla)\vec{U}]\cdot [n\times \vec{U}]=0.
\end{align}
For clarity, we collect the modified NBC here
\begin{align}
& \vec{U} \cdot \hat{n}=0,\\
&  \vec{U} \cdot [(\hat{n}\cdot \nabla)\vec{U}]=-\frac{1}{b_w} \vec{U} \cdot \vec{U},\\
\label{APBNC3}
& [(\hat{n}\cdot \nabla)\vec{U}]\cdot [\hat{n}\times \vec{U}]=0.
\end{align}

\section{}
Before we resolve the $K$, we need to decompose the $\nabla \times U$ by two vectors. One vector is along the $\hat{n}\times \hat{\tau}$. The other vector is in the $y,z$ plan. The decomposition has been shown in the eq.(\ref{nablaU}). Now we simplify the numerators of the equation. The factors of the terms are simplified as the following.
\begin{align}
\label{ACV1}
&(\nabla \times \vec{U})\cdot (\hat{n}\times \vec{U})=n_jU_k \frac{\partial U_k}{\partial x_j}-n_kU_j \frac{\partial U_k}{\partial x_j} \nonumber \\
&=\vec{U}\cdot [(\hat{n}\cdot \nabla)\vec{U}]-\hat{n}\cdot [(\vec{U}\cdot \nabla)\vec{U}]\nonumber \\
&=-\frac{1}{b_w} \vec{U}\cdot \vec{U}-\hat{n}\cdot [(\vec{U}\cdot \nabla)\vec{U}].
\end{align}
Thus, for the eq.(\ref{ACV1}), we have
\begin{align}
&\hat{n}\times [(\nabla \times \vec{U})\cdot (\hat{n} \times \vec{U})](\hat{n} \times \vec{U})\nonumber \\
&=[-\frac{1}{b_w} \vec{U}\cdot \vec{U}-\hat{n}\cdot [(\vec{U}\cdot \nabla)\vec{U}]](\hat{n}\times \hat{n} \times \vec{U}).
\end{align}
And
\begin{align}
\label{ACV2}
&(\hat{n}\times \vec{U})\times (\nabla \times \vec{U})\times (\hat{n}\times \vec{U})\nonumber \\
&=\epsilon_{bef}i_an_eU_f(n_cU_a\frac{\partial U_c}{\partial x_b}+n_aU_c\frac{\partial U_b}{\partial x_c}\nonumber \\
&~~~~-n_aU_c\frac{\partial U_c}{\partial x_b}-n_cU_a\frac{\partial U_b}{\partial x_c})\nonumber \\
&=\vec{U} \hat{n}\cdot [(\hat{n}\times \vec{U})\cdot \nabla]\vec{U}+\hat{n}(\hat{n}\times \vec{U})\cdot [(\vec{U}\cdot \nabla)\vec{U}]\nonumber \\
&~~~~-\hat{n} \vec{U}\cdot [(\hat{n}\times \vec{U})\cdot \nabla]\vec{U}-\vec{U}[(\hat{n}\cdot \nabla)\vec{U}\cdot (\hat{n}\times \vec{U})].
\end{align}
The last term of the above equation is dropped off by using the eq.(\ref{APBNC3}). For the eq.(\ref{ACV2}), we have
\begin{align}
&\hat{n}\times (\hat{n}\times \vec{U})\times (\nabla \times \vec{U})\times (\hat{n}\times \vec{U})\nonumber \\
&=(\hat{n}\times \vec{U}) [\hat{n}\cdot [(\hat{n}\times \vec{U})\cdot \nabla]\vec{U}],
\end{align}
By using the equality of $\hat{n}\times \hat{n}=0$. In this way, the $\hat{n}\times \nabla \times \vec{U}$ can be split into three terms by
\begin{align}
&T_1=\frac{- \vec{U}\cdot \vec{U}(\hat{n} \times \hat{n} \times \vec{U})}{b_w( \hat{n}\times \vec{U})\cdot ( \hat{n}\times \vec{U})},\nonumber\\
&T_2=\frac{-\hat{n}\cdot ((\vec{U}\cdot \nabla)\vec{U})(\hat{n} \times \hat{n} \times \vec{U})}{( \hat{n}\times \vec{U})\cdot ( \hat{n}\times \vec{U})},\nonumber\\
&T_3=\frac{(\hat{n} \times \vec{U}) [\hat{n}\cdot ((\hat{n}\times \vec{U})\cdot \nabla)\vec{U}]}{(\hat{n}\times \vec{U})\cdot (\hat{n}\times \vec{U})}.\nonumber 
\end{align}


\begin{thebibliography}{9}
\bibitem{Eijkel}
J. C. T. Eijkel and A. van den Berg, \textit{Nanofluidics: what is it and what can we expect from it?}  Microfluid Nanofluid., {\bf 1},249(2005)
\bibitem{Bocquet1}
Lyderic Bocquet and Elisabeth Charlaix, \textit{Nanofluidics, from bulk to interfaces}, Chem. Soc. Rev. {\bf 39},1073(2010)
\bibitem{Siwy}
Z. Siwy and A. Fulinski, \textit{Fabrication of a Synthetic Nanopore Ion Pump}, Phys. Rev. Lett., {\bf 89},198103(2002)
\bibitem{Karnik1}
R. Karnik, R. Fan, M. Yue, D. Li, P. Yang and A. Majumdar, \textit{Electrostatic Control of Ions and Molecules in Nanofluidic Transistors}, Nano Lett., {\bf 5},943(2005)
\bibitem{Schasfoort}
R. B. M. Schasfoort, S. Schlautmann, J. Hendrikse and A. van den Berg, \textit{Field-Effect Flow Control for Microfabricated Fluidic Networks},Science, {\bf 286},942(1999)
\bibitem{Karnik2}
R. Karnik, C. Duan, K. Castelino, H. Daiguji and A. Majumdarn, \textit{Rectification of Ionic Current in a Nanofluidic Diode}, Nano Lett., {\bf 7},547(2007)
\bibitem{Chan}
D. Y. Chan and R. G. Horn, \textit{The drainage of thin liquid films between solid surfaces}, J. Chem. Phys., {\bf 83},5311(1985)
\bibitem{Raviv}
U. Raviv and J. Klein, \textit{Fluidity of Bound Hydration Layers}, Science, {\bf 297},1540(2002)
\bibitem{Bocquet2}
L. Bocquet and J. -L. Barrar,  \textit{Flow boundary conditions from nano- to micro-scales}, Soft matter, {\bf 3},685(2007)
\bibitem{Maali}
A. Maali, T. Cohen-Bouhacina, G. Couturier and J. -P. Aime, \textit{Oscillatory Dissipation of a Simple Confined Liquid}, Phys. Rev. Lett., {\bf 96},086105(2006)
\bibitem{Georges}
J. -M. Georges, S. Millot, J. -L. Loubet and A. Tonck, \textit{Drainage of thin liquid films between relatively smooth surfaces}, J. Chem. Phys., {\bf 98},7345(1993)
\bibitem{Becker}
T. Becker and F. Mugele,  \textit{Nanofluidics: Viscous Dissipation in Layered Liquid Films}, Phys. Rev. Lett., {\bf 91},166104(2003)
\bibitem{Leng}
Y. Leng and P. T. Cummings, \textit{Fluidity of Hydration Layers Nanoconfined between Mica Surfaces}, Phys. Rev. Lett., {\bf 94},026101(2005)
\bibitem{Li}
T. -D. Li, J. Gao, R. Szoszkiewicz, U. Landman and E. Riedo, \textit{Structured and viscous water in subnanometer gaps}, Phys. Rev. B., {\bf 75},115415(2007)
\bibitem{Batchelor}
G. K. Batchelor, \textit{an Introduction to Fluid Dynamics} (Cambridge Univeristy Press, Cambridge, 1967)
\bibitem{Bocquet3}
L. Bocquet and J. -L. Barrat, \textit{Hydrodynamic boundary conditions, correlation functions, and Kubo relations for confined fluids}, Phys. Rev. E., {\bf 49},3079(1994)
\bibitem{Pit}
R. Pit, H. Hervet and L. Leger, \textit{Direct Experimental Evidence of Slip in Hexadecane: Solid Interfaces}, Phys. Rev. Lett., {\bf 85},980(2000)
\bibitem{Cottin1}
C. Cottin-Bizonne, C. Barentin, E. Charlaix, L. Bocquet and J. -L. Barrat, \textit{Dynamics of simple liquids at heterogeneous surfaces: Molecular-dynamics simulations and hydrodynamic description}, Eur. Phys. J. E, {\bf 15},427(2004)
\bibitem{Priezjev}
N. V. Priezjev and S. M. Troian, \textit{Molecular Origin and Dynamic Behavior of Slip in Sheared Polymer Films}, Phys. Rev. Lett., {\bf 92},018302(2004)
\bibitem{Jabbarzadeh}
A. Jabbarzadeh, P. Harrowell and R. I. Tanner, \textit{Low friction lubrication between amorphous walls: unraveling the contributions of surface roughness and in-plane disorder}, J. Phys. Chem., {\bf 125},034703(2006)
\bibitem{Thompson}
P. A. Thompson and M. O. Robbins, \textit{Shear flow near solids: Epitaxial order and flow boundary conditions}, Phys. Rev. A, {\bf 41},6830(1990)
\bibitem{Cottin2}
C. Cottin-Bizonne, B. Cross, A. Steinberger and E. Charlaix, \textit{Boundary Slip on Smooth Hydrophobic Surfaces: Intrinsic Effects and Possible Artifacts}, Phys. Rev. Lett., {\bf 94},056102(2005)
\bibitem{Cieplak}
M. Cieplak, J. Koplik and J. R. Banavar, \textit{Boundary Conditions at a Fluid-Solid Interface}, Phys. Rev. Lett., {\bf 86},803(2001)
\bibitem{Zhu}
Y. Zhu and S. Granick, \textit{Rate-Dependent Slip of Newtonian Liquid at Smooth Surfaces}, Phys. Rev. Lett., {\bf 87},096105(2001)
\bibitem{Liu}
Tingyi Liu, Chang-Jin Kim, \textit{Turning a surface superrepellent even to completely wetting liquids}, Science , {\bf 346},1096(2014)
\bibitem{Ou1}
J. Ou, B. Perot and J. P. Rothstein, \textit{Laminar drag reduction in microchannels using ultrahydrophobic surfaces}, Phys. Fluids, {\bf 16},4635(2004)
\bibitem{Ou2}
J. Ou and J. P. Rothstein, \textit{Direct velocity measurements of the flow past drag-reducing ultrahydrophobic surfaces}, Phys. Fluids, {\bf 17},103603(2005)
\bibitem{Choi1}
C. H. Choi, U. Ulmanella, J. Kim, C. M. Ho and C. J. Kim, \textit{Effective slip and friction reduction in nanograted superhydrophobic microchannels}, Phys. Fluids, {\bf 18},087105(2006)
\bibitem{Choi2}
C. H. Choi and C. J. Kim, \textit{Large Slip of Aqueous Liquid Flow over a Nanoengineered Superhydrophobic Surface}, Phys. Rev. Lett., {\bf 96},066001(2006)
\bibitem{Callies}
M. Callies and D. Quere, \textit{On water repellency}, Soft Matter, {\bf 1}, 55 (2005)
\bibitem{Truesdell}
R. Truesdell, A. Mammoli, P. Vorobieff, F. van Swol and
C. J. Brinker, \textit{Drag reduction on a patterned superhydrophobic surface}, Phys. Rev. Lett., {\bf 97}, 044504(2006)
\bibitem{Bocquet4}
L. Bocquet, P. Tabeling and S. Manneville, \textit{Comment on "Large slip of aqueous liquid flow over a nanoengineered superhydrophobic surface"}, Phys. Rev. Lett.,{\bf 97}, 109601( 2006)
\bibitem{Joseph}
P. Joseph, C. Cottin-Bizonne, C. Y. J.-M. Benoit, C. Journet,
P. Tabeling and L. Bocquet,\textit{Slippage of water past superhydrophobic carbon nanotube forests in microchannels}, Phys. Rev. Lett., {\bf 97}, 156104(2006)
\bibitem{Lauga}
E. Lauga and H. A. Stone, \textit{Effective slip in pressure-driven Stokes flow }, J. Fluid Mech., {\bf 489}, 55(2003)
\bibitem{Richardson}
S. Richardson, \textit{On the no-slip boundary condition}, J. Fluid Mech., {\bf 59}, 707(1973)
\bibitem{Cottin4}
C. Cottin-Bizonne, J.-L. Barrat, L. Bocquet and E. Charlaix, \textit{Low-friction flows of liquid at nanopatterned interfaces}, Nat.
Mater., {\bf 2}, 237(2003)
\bibitem{Hu}
H. W. Hu, G. A. Carson, and S. Granick, \textit{Relaxation time of confined liquids under shear}, Phys. Rev. Lett., {\bf 66},2758(1991)
\bibitem{Israelachvili}
J. N. Israelachvili, P. M. McGuiggan, and A. M. Homola, \textit{Dynamic Properties of Molecularly Thin Liquid Films}, Science, {\bf 240},189(1988)
\bibitem{Chen}
Shuyu Chen, Han Wang, Tiezheng Qian, and Ping Sheng, \textit{Determining hydrodynamic boundary conditions from equilibrium fluctuations}, Phys. Rev. E, {\bf 92},043007(2015)
\bibitem{Thompson1}
P. A. Thompson and S. M. Troian, \textit{A general boundary condition for liquid flow at solid surfaces}, Nature, {\bf 389},360(1997)
\bibitem{Thompson2}
P. A. Thompson, G. S. Grest and M. O. Robbins, \textit{Phase transitions and universal dynamics in confined films}, Phys. Rev. Lett., {\bf 68},3448(1992)
\bibitem{Smith}
E. D. Smith, M. O. Robbins and M. Cieplak, \textit{Friction on adsorbed monolayers}, Phys. Rev. B, {\bf 54},8252(1996)
\bibitem{Walz}
T. Walz, B. L. Smith, M. L. Zeidel, A. Engel and P. Agre, \textit{Biologically active two-dimensional crystals of aquaporin CHIP}, J. Biol. Chem., {\bf 269},1583(1994)

\end{thebibliography}

\end{document}